\documentclass[useAMS,usenatbib,usegraphicx]{mn2e}

\title[Stability, chaos and entrapment of stars in very wide pairs]{Stability, chaos and entrapment of stars in very wide pairs}
\author[Valeri V. Makarov]{Valeri V. Makarov$^{1}$\thanks{E-mail:
vvm@usno.navy.mil}\\
$^{1}$US Naval Observatory, 3450 Massachusetts Ave NW, Washington DC 20392-5420, USA}
\begin{document}

\date{Accepted . Received ; in original form }

\pagerange{\pageref{firstpage}--\pageref{lastpage}} \pubyear{2002}

\maketitle

\label{firstpage}

\begin{abstract}
The relative motion of  stars and other celestial objects in very wide pairs,  separated by distances of the order of 1 pc, is strongly influenced by the tidal gravitational potential of the Galaxy. The Coriolis component of the horizontal tidal force in the rotating reference frame tends to disrupt such marginally bound pairs. However, even extremely wide pairs of bodies can be bound over intervals of time comparable to the Hubble time, under appropriate initial conditions. Here we show that for arbitrary chosen initial coordinates of a pair of stars, there exists a volume of the space of initial velocity components where the orbits remain bound in the planar tidal field for longer than 10 Gyr, even though the initial separation is well outside the Jacobi radius. The boundary of this phase space of stable orbits is fractal, and the motion at the boundary conditions is clearly chaotic. We found that the pairs may remain confined for several Gyr, and then suddenly disintegrate due to a particularly close rendezvous. By reversing such long-term stable orbits, we find that entrapment of unrelated stars into wide pairs is possible, but should be quite rare. Careful analysis of precision astrometry surveys revealed that extremely wide pairs of stars are present in significant numbers in the Galaxy. These results are expected to help discriminating the cases of genuine binarity and chance entrapment, and to make inroads in testing the limits of Newtonian gravitation.\end{abstract}


\section{Introduction}

The Galactic tidal force becomes comparable to the force of gravitational attraction of
widely separated double stars at distances close to the Jacobi radius, 
which approximately equals 1.8 parsecs for twin stars of solar mass. Beyond the Jacobi
radius, the planar component of the tidal force may cause a relatively rapid disruption
of a wide pair. Stars emerging from disrupted binary systems or stellar associations are stretched out by the Galactic potential into narrow streams in the direction of Galactic rotation, a process, which is well described by the epicycle approximation \citep{maol}. This does not mean that all pairs of stars separated by distances greater than the Jacobi radius will be disrupted by the tidal force. The Coriolis component of force acting on a pair of stars in the co-rotating reference frame centered on the primary (i.e., larger mass) companion is proportional to the instantaneous relative velocity of the secondary companion
\citep{jia}. Under appropriate initial conditions, the Coriolis force can confine the relative motion of two stars to a finite volume of the phase space, even if their mutual distance is far greater than the Jacobi radius. The resulting trajectories are similar to regular orbits, in that they have a periodic character and the separation never exceeds a certain limit within the Hubble time.

In this paper, we present and discuss the results of a large number of numerical integration of
pairs of stars in the rotating, noninertial reference frame, performed with the J. Chambers' Mercury code,
appropriately modified for this case (see Appendix).
The aim of these numerical simulations is to demonstrate that a certain area of the phase space of
initial parameters exists well outside the Jacobi radius, where the resulting orbits are regular,
bound and stable. By retracing long-term stable orbits at the boundary of the stable area, we prove
that accidental capture of two passing stars into a stable double system is possible in principle.

\section{Stability of extremely wide double stars}
\begin{figure}
\includegraphics[width=85mm]{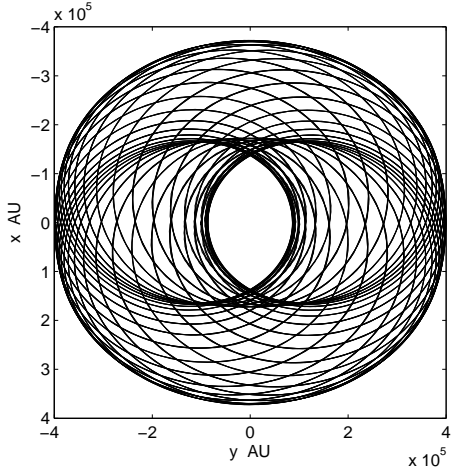}
\caption{Trajectory of a 0.72 $M_{\sun}$ companion around a 1 $M_{\sun}$ star in the plane of the Galaxy, integrated over 10 Gyr. The $x$ axis in this co-rotating reference frame is pointing toward the Galactic center and the $y$ axis is in the direction of Galactic rotation. The primary component is at the coordinate origin.}
\label{stab.fig}
\end{figure}
Fig. \ref{stab.fig}a shows an example of a stable orbit for a hypothetical pair of stars with a total mass of 1.72 solar masses, traveling on a circular orbit at the orbital radius of the Sun around the Galaxy.   The primary star is at the origin of the reference frame ($x=0$, $y=0$), and the initial phase space parameters of the companion are $x_0=0$, $y_0=-4\cdot10^5$  AU, $\dot x_0=4.8\cdot10^{-5}$ AU d$^{-1}$, $\dot y_0=0$. The companion, moving in shifting double loops around the primary for 10 Gyr (which was our integration time) never wanders away by more than the initial separation. The Jacobi radius for this case is equal to 
$3.55\cdot10^5$ AU, assuming the values for the OortÕs constants determined from the Hipparcos data
\citep{mamu}. Throughout this paper, the $x$ axis is directed toward the Galactic center, the $y$ axis toward the direction of Galactic rotation, and our computations are restricted to the planar case, as explained in Appendix. 

Such long-term stable orbits at extreme separations are only possible when the orbital 
angular momentum is opposite to the Galactic rotation, i.e., when the relative motion is retrograde. 
As seen from the North Galactic pole, the orbital motion is always counterclockwise in stable 
pairs. This important property has been deduced for the classical restricted three-body problem 
\citep{hen} and, more recently, for regular orbits of stars around open clusters in the Galactic tidal potential \citep{fuk}. 
A co-rotating pair of stars beyond the Jacobi radius is disrupted by the Coriolis acceleration 
well before it completes a full revolution in the Galaxy (approximately, 230 Myr). 
Thus, the chances of survival of very wide double systems are defined by initial conditions, 
barring external perturbations from other field stars. By integrating numerous trajectories at 
a step of 0.1 AU d$^{-1}$ in $\dot x_0$  and $\dot y_0$  for the previously used initial 
coordinates  $x_0=0$, $y_0=-4\cdot10^5$  AU, we found that the area of stable orbits is approximately triangular in shape defined by the vertices at 
$(\dot x_0,\dot y_0)=\{(-4.74,0),(+4.74,0),(0,7.0)\}\cdot 10^{-5}$ AU d$^{-1}$.  
Most of the trajectories with initial velocities inside this triangle are bound in the 
long term (over 10 Gyr or longer). Most of the trajectories outside this triangle are not bound, 
with the distance rapidly increasing beyond the initial separation. 

\section{Chaos and entrapment}
\begin{figure}
\includegraphics[width=75mm]{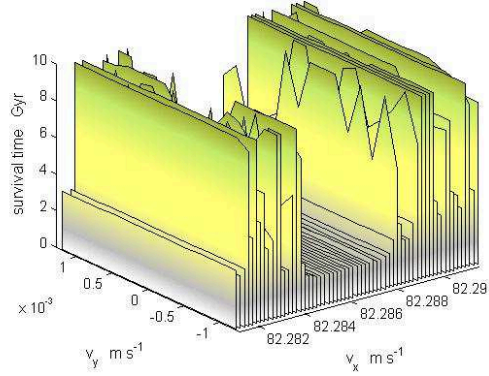}
\caption{Survival time of pairs of stars initially separated by $400\,000$ AU computed with a step 
of 0.17 mm s$^{-1}$ in initial velocity components. A tiny variation in phase space parameters 
may cause a drastic change in stability at this transitional region separating bound and unbound trajectories.}
\label{fract.fig}
\end{figure}

The boundary of the phase space of stable orbits assisted by the Galactic tidal field is fractal, 
as established numerically by performing integrations at a small step in the initial conditions. 
In the vicinity of this boundary, a tiny change in the initial velocity, for example, may alter 
the fate of the stellar pair, resulting in orbits that are bound for longer than 10 Gyr, or ones that 
fall apart within 100 Myr. Fig. \ref{fract.fig} shows the survival time of pairs with the initial coordinates at  
$x_0=0$, $y_0=-4\cdot10^5$ AU, and initial velocities varied by small amounts around  $\dot x_0=4.7394\cdot
10^{-5}$ AU d$^{-1}$, $\dot y_0=0$. Sudden changes in survival time take place on scales as small 
as 0.17 mm s$^{-1}$ in either dimension of the velocity space.  In a separate experiment, we obtained 
a stable orbit over 10 Gyr at $\dot x_0=4.7398\cdot10^{-5}$ AU d$^{-1}$, an orbit which disintegrated 
in 3.26 Gyr at $\dot x_0=4.73981\cdot10^{-5}$ AU d$^{-1}$, and, most remarkably, an orbit which 
fell apart after 9 Gyr at $\dot x_0=4.7398000004\cdot10^{-5}$ AU d$^{-1}$. The machine precision 
of these computations is approximately one part in $10^{15}$.

The motion of stars with initial parameters at the boundary of the space of stable orbits is strongly 
chaotic. Using the method to estimate the Lyapunov time suggested for the outer Solar System
\citep{hay07,hay08}, two sibling trajectories were integrated at $x_0=0$, $y_0=-4\cdot10^5$ AU, 
$\dot x_0=4.7394\cdot10^{-5}$ AU d$^{-1}$, $\dot y_0=0$ by changing the initial value 
$\dot x_0$  by one part in $10^{11}$.  The spatial separation between the siblings grew exponentially 
over the first few Gyr, with an overlaid oscillation of 0.72 Gyr period. The estimated Lyapunov time 
was 0.56 Gyr, which is approximately equal to the duration of two Galactic revolutions. Both orbits 
were found stable over 10 Gyr.

\begin{figure}
\includegraphics[width=75mm]{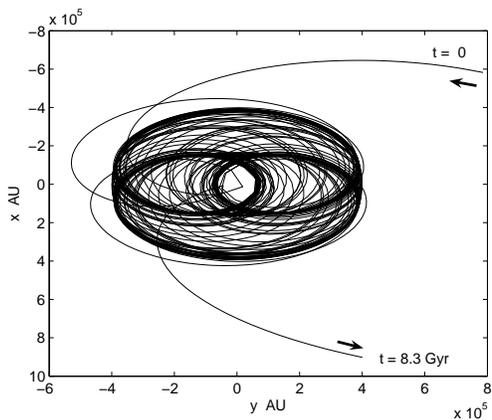}
\caption{Trajectory of a 0.72 $M_{\sun}$ companion around a 1 $M_{\sun}$ star in the plane of the Galaxy, entrapped into a bound pair for 8.3 Gyr by the Galactic tidal force. Both the capture and the ejection 
of the companion were achieved through gravitational maneuvers in the vicinity of the primary component.}
\label{capt.fig}
\end{figure}

The long-term stable obit at $\dot x_0=4.7398000004\cdot10^{-5}$ AU d$^{-1}$ , discussed above, 
is of special interest for the problem of capture of random passing stars into double systems. 
This pair was bound for 9 Gyr, after which it suddenly fell apart when the secondary component wandered 
too close to the primary and was accelerated by its gravitational pull. Once the pair is torn 
apart, the distance between the components begins to grow rapidly. We can construct a long-term stable 
pair by taking any of the points on the outgoing trajectory, reversing the time and performing a 
reflection in the four-dimensional phase space. The reflection is necessary, because only retrograde 
trajectories are stable. A stable trajectory obtained this way for a pair of initially unbound stars 
is shown in Fig. \ref{capt.fig}. A chance fly-by star was decelerated by the tidal force, went 
through a close periastron where it lost more kinetic energy and started to orbit the primary 
on the characteristic double loops we have seen in Fig. \ref{stab.fig}.  After 
$\sim$8 Gyr of quasi-regular motion, the companion wandered too close to the primary and was ejected 
again. Obviously, the original trajectory, which was bound for $\sim$9 Gyr, could not be exactly 
reproduced in this experiment. The chaotic motion and the fractal character of the boundary between 
long-term and short-term trajectories make the machine precision (about 16 significant digits) of our 
calculations insufficient to match two orbits over such a long time. Thus, accidental entrapment 
of passing stars into stable pairs is possible in principal, but should occur extremely seldom 
because of the small area of suitable phase space parameters. Furthermore, any perturbation from 
a third passing body will drastically change the fate of the system, in most cases resulting in a 
relatively rapid breakup. Occasionally, the disturbance from a chance fly-by will drive an initially 
unstable pair into the space of stable trajectories, but the probability of such events at separations 
above the Jacobi radius is very small. This entrapment mechanism is similar to the chaos-assisted
capture of irregular moons by the Sun-Jupiter system, which can only happen in the thin layer
of chaotic motion separating the regions of scattering and stability \citep{ast}.

\section{Discussion}
It follows from these considerations that the entrapment of random stars into extremely wide systems 
has low probability and that such associations should be short-lived. Yet, a growing body of 
observational evidence indicates that pairs of stars with separations up to 0.1 Ð- 1 pc are abundant 
even in the close neighborhood of the Sun \citep{mak,cab,sha}.  Some of these pairs may be the 
remnants of disintegrating open clusters or OB associations \citep{kou}, but not all of the components are 
sufficiently young. If verified by accurate astrometric observations, the existence of only a 
few pairs of stars with separations greater than the Jacobi radius in the Solar neighborhood 
will challenge the classical theory of gravitation in the domain of weak forces. Any modification 
of Newtonian dynamics in the weak-field regime that boosts the gravitational acceleration at large 
distances will drastically increase both the probability of entrapment of random pairs of field stars 
and the survival time of very wide binaries. 
Even the nearest known stellar system to us, the 
Alpha Centauri AB and the remote Proxima Centauri companion, may pose a problem for Newtonian 
dynamics because of the relatively large difference in velocity \citep{ano, wer}. \citet{bee} suggested that
the apparent stability of this triple system may provide a crucial test for the MOND \citep{mil},
or the theory of a coupling between the scalar curvature and the matter Lagrangian density \citep{orf},
where an additional force can widen the allowable limit on the orbital acceleration.  
For the case studied in this paper ($M_{\rm tot}=1.72 M_{\sun}$, $y_0=-4\cdot10^5$  AU), the
Newtonian gravitational acceleration is much smaller than the threshold acceleration of
MOND ($a_0\approx 1.2\cdot 10^{-10}$ m s$^{-2}$). Using the asymptotic expression for acceleration
of the companion star, $a=\sqrt{G\,M_{\rm tot}\,a_0}\,r^{-1}$, we estimate the distance of the Lagrange
points to be
\begin{equation}
r_J=\left(\frac{\sqrt{G\,M_{\rm tot}\,a_0}}{4A\Omega}\right)^\frac{1}{2}
\end{equation}
which for this pair of stars is $r_J=2.2\cdot 10^6$ AU, or $10.5$ pc. The Jacobi radius in MOND is 6.2 times larger
than the Newtonian Jacobi radius, and the area of stable orbits is roughly 38 times greater in the $(x,y)$ plane.
Very accurate
radial velocity measurements of Proxima Cen would be required for a conclusive test \citep{bee11}. 
Very wide pairs of nearby stars, 
due to even stricter criteria of survival, may provide the most sensitive experimental data for 
testing the theories of gravitation in the weak-field regime. 

\section*{Acknowledgments}
I thank the USNO Editorial Board for helpful suggestions and
a critical reading of the original version of the
paper.

\appendix
\section{Galactic tidal force}
The planar components of the Galactic tidal acceleration are accurately approximated by the epicycle model for all objects moving on nearly circular orbits at the solar radius\citep{jia,kin,maol}:
\begin{eqnarray}
\ddot x&=&4A\Omega x-2\Omega \dot y \nonumber\\
\ddot y&=&2\Omega\dot x \nonumber
\end{eqnarray}
where $A$ is the OortÕs constant, and $\Omega$ is the rotation frequency. 
The vertical component of acceleration, to a good approximation for small vertical velocities at the plane, 
is harmonic\citep{maol}, 
and, therefore, tends to compress very wide binaries. It is sufficient to consider the stability 
problem restricted to the planar case, ignoring the vertical dimension. All integrations were 
performed with the well-tested Mercury code \citep{cha} by adding the tidal acceleration components 
in the user-defined external acceleration subroutine. In the units adopted in the Mercury code, 
the assumed parameters were 
$4 A \Omega=1.19\cdot 10^{-20}$  d$^{-2}$ and  $2\Omega=1.5\cdot10^{-10}$ d$^{-1}$. 
We used the Bulirsch-Stoer option of integration and set the integration step to 200 d. The companion 
was considered ejected when the distance from the primary exceeded 900 000 AU. 

\bsp

\label{lastpage}

\end{document}